\documentclass{PoS}
\usepackage{graphicx}
\usepackage{graphics}
\usepackage{colordvi}
\usepackage{multirow}
\usepackage{multicol}
\usepackage{amssymb,amsmath}
\def\Eq#1{Eq.~(\ref{#1})}
\def\Ref#1{Ref.~\cite{#1}}
\def\Sec#1{Section~\ref{#1}}
\title{The spectrum of closed loops of fundamental flux in $D=3+1$ $SU(N)$ gauge theories.}

\ShortTitle{Spectrum of closed fundamental flux tubes in $D=3+1$}

\author{\speaker{Andreas Athenodorou} \\ NIC, DESY, \\ Platanenallee 6, 15738 Zeuthen, Germany \\ and \\ Rudolf Peierls Centre for Theoretical Physics \\
        University of Oxford \\
        E-mail: \email{andreas.athenodorou@desy.de}}

\author{Barak Bringoltz \\ Physics Department\\
        University of Washington, Seattle, WA 98195-1560, USA\\
        E-mail: \email{barak@phys.washington.edu}
}

\author{Michael Teper  \\ Rudolf Peierls Centre for Theoretical Physics \\
        University of Oxford\\
    E-mail: \email{m.teper1@physics.ox.ac.uk}}

\abstract{
We study the spectrum of closed flux tubes in four dimensional $SU(N)$ gauge theories. We do so by calculating the energies of the low lying states with the  variational technique (whose basis consists of about $\sim 700$ operators). We study states of different values of angular momentum, transversal  parity, longitudinal parity, and longitudinal momentum, and compare the results with effective string theories (ESTs) such as the Nambu-Goto (NG) model. Most of our states agree very well with the Nambu-Goto predictions and since most of our flux-tubes' lengths are outside the radius of convergence of the ESTs, then for some states it is only the NG that predicts the spectrum well. This strongly suggests that the ESTs can be re-summed. Nonetheless, there are a few states (all with negative parity and in the same representation of the lattice rotation group) that exhibit large deviations from the NG predictions; these deviations might provide clues to the nature of the effective string theory describing the large-$N$ QCD string.
}

\FullConference{The XXVII International Symposium on Lattice Field Theory\\
         July 25-31 2009\\
         Beijing, China}

\begin{document}
\section{Introduction}
\vskip -0.25cm
We have recently shown \cite{ABM} that the closed flux tube spectrum in $D=2+1$ $SU(N)$ gauge theories can be well-approximated by the Nambu-Goto (NG) free string 
in flat space-time. In particular, such agreement was observed for flux tube lengths that are comparable to the width 
of the flux. Thus, the flux-tube approximately behaves like a fundamental string even when, naively, it looks more like a fat blob than a thin string.

It is interesting to know how the closed flux tube spectrum in $D=3+1$ $SU(N)$ gauge theories behaves and how this compares to NG and other effective string theories. A pioneering attempt to calculate the closed flux tube spectrum in $D=3+1$ $SU(N)$ gauge theories has been reported in~\cite{Kuti1}, which investigated the spectrum for $N=3$ and for a single, rather coarse, spatial lattice spacing of $a\approx 0.22 \mathrm{fm}$. For earlier works that focused on the ground state of the closed string see \cite{old} and its citations.
In this paper we present results from three calculations for the closed flux tube spectrum. Two for 
$N=3$ at $\beta=6.0625$ ($a\simeq 0.09 {\rm fm}$) and at $6.338$ ($a\simeq 0.06 {\rm fm}$), and one for $N=5$ at $\beta=17.630$ ($a\simeq 0.09 {\rm fm}$). In a further publication we will present high statistics measurements of the ground state for $SU(3)$ and $SU(6)$ at $a\approx 0.09$fm. All our $N>3$ calculations are focused only on flux-tubes that carry a single unit of electric flux.

\vspace{-0.5cm}
\section{Setup of the lattice calculation}
\vskip -0.25cm
We define the $SU(N)$ gauge theory on a four-dimensional Euclidean space-time lattice which is  compactified along all directions with 
$L_{\parallel} \times L_{\perp_1} \times L_{\perp_2} \times L_{T}$ sites. The length of the flux tube is equal to $L_{\parallel}$, while $L_{\perp_1}$, 
$L_{\perp_2}$ and $L_{T}$ were chosen to be large enough so to avoid finite volume effects. To extract the 
flux tube spectrum we perform Monte-Carlo simulations using the standard Wilson plaquette action, 
%\begin{eqnarray}
$S=\sum_{\square}\beta \left[ 1-\frac{1}{N}{\rm Re}{\rm Tr}(U_{\square}) \right]$, with $\beta=\frac{2N}{g^2(a)}$,
%\label{plaquetteaction}
%\end{eqnarray}
and in order to keep the value of the lattice spacing $a$ approximately  fixed for different values of $N$ we keep
the 't Hooft coupling $\lambda(a)=N g^2(a)$ approximately fixed, so that $\beta \propto N^2$. The simulation algorithm we use combines standard heat-bath and over-relaxation steps in the ratio 1:4; these are implemented by updating $SU(2)$ subgroups using the Cabibbo-Marinari algorithm. To measure the spectrum of energies we use the variational technique 
(e.g.~see \Ref{variational} and its references).

The closed flux tube states in $D=3+1$ can be classified by the irreducible representations of the two-dimensional lattice cubic symmetry which we denote by $C_4$.\footnote{Since the global rotational symmetries of a flux-tube wound around a compact direction are those of a two-dimensional space.} It is a subgroup of $O(2)$ that corresponds to rotations by integer multiples of $\pi/2$ around the tube axis. This makes values of angular momenta that differ by an integer multiple of four indistinguishable on the lattice, and factorizes the Hilbert space into four sectors: $J_{{\rm mod}\, 4}=0$, $J_{{\rm mod}\, 4}=\pm 1$,  $J_{{\rm mod}\, 4}=2$. An additional useful quantum number is the parity $P_\perp$ which is associated with reflections around the axis $\hat{\perp}_1$. Such parity flips the sign of $J$ and so we can choose a basis in which states are characterised by their value of $J$ (which can be of either sign), or by their value of $|J|$ and  $P_\perp$. In our calculations we use the latter choice. While in the continuum states of nonzero $J$ are parity degenerate, on the lattice this is exactly true only for the odd $J$ sector. This means we can denote our states by the $5$ irreducible representations $A_{1,2}, E, B_{1,2}$ of $C_4$ whose $J$ and $P_\perp$ assignments are: $\left\{ A_1: \, |J_{{\rm mod}4}|=0, P_\perp=+\right\}, \left\{A_2:  \, |J_{{\rm mod}4}|=0, P_\perp=-\right\}$, $\left\{ E: |J_{{\rm mod}4}|=1, P_\perp=\pm\right\},$ $\left\{ B_1: \, |J_{{\rm mod}4}|=2, P_\perp=+\right\}$, and 
$\left\{B_2:  \, |J_{{\rm mod}4}|=2, P_\perp=-\right\}$.
All the representations of $C_4$  are one-dimensional except for $E$ which is two-dimensional. 

Two additional useful quantum numbers include the longitudinal momentum $p_{||}$ carried by the flux-tube along its axis (which is quantized in the form $p_{||}=2\pi q/L_{||}; q\in Z$) and the parity $P_{||}$ with respect to reflections across the string midpoint. Since $P_{||}$ and $p_{||}$ do not commute, we can use both to  simultaneously characterise a state only when $q=0$. Also, since the energy does not depend on the sign of $q$, we only focused on those with $q\ge 0$.

The operators we construct have shapes that lead to certain values of $J,P_\perp,P_{||},$ and $q$. This is achieved by choosing a 
linear combination of Polyakov loops whose paths consist of  various  transverse deformations and various smearing and blocking levels (again see \cite{variational}). All the paths used for the construction of the operators are presented in Table~\ref{operatorstable} and all together form a basis of around 700 operators. Let us show how to  construct an operator with a certain value of $J_{{\rm mod}4}$: begin with the operator $\phi_{\alpha}$ that has a deformation extending in angle $\alpha$ within the plane of transverse directions. We can construct an operator $\phi(J)$ that belongs to a specific representation of $C_{4}$ by using the formula: 
%\vspace{-0.25cm} 
%\begin{eqnarray}
$\phi(J)= \sum_{n=1,2,3,4} e^{iJ n \frac{\pi}{2}} \phi_{n \frac{\pi}{2}}.$
%\end{eqnarray}
%\vspace{-0.65cm} 
It is straight-forward to show that $\phi(0)$ belongs to either $A_1$ or $A_2$ (depending on its value of $P_\perp$), that $\phi(1)$ belongs to $E$, and that $\phi(2)$ belongs to $B_1$ or $B_2$. The projection onto  certain values of $P_\perp$ and $P_{||}$ is demonstrated pictorially in  \Eq{A12} for an operator of $J_{{\rm mod}4}=0$.
%\vspace{-0.15cm} 
\begin{eqnarray}
\hspace{-0.25cm} \phi= {\rm Tr}\left[ \parbox{12.5cm}{\rotatebox{0}{\includegraphics[width=12.5cm]{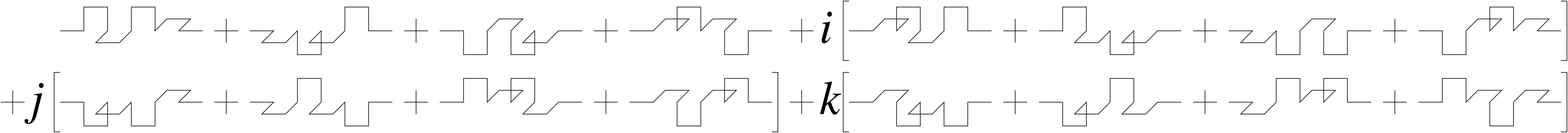}}} \ \right] \label{A12}
\end{eqnarray}
%\vspace{-0.5cm} 
%
%\noindent
If $i=j=k=+1$ then the operator $\phi$ projects onto $\{ A_1, P_{||}=+ \}$, if $i=+1,j=k=-1$ then it projects onto 
$\{ A_2, P_{||}=+ \}$, if $i=-1,j=+1,k=-1$ it projects onto $\{ A_1, P_{||}=- \}$ and finally, if $i=j=-1,k=+1$, it projects onto 
$\{ A_2, P_{||}=- \}$.
  \begin{table}[htp]
\centering{\scalebox{1.125}{
\begin{tabular}[c]{|c||c||c||c||c||c||c||c||c||c||c||c||c|} \hline \hline
1 & 2 & 3 & 4 & 5 & 6 & 7 & 8 & 9 & 10 & 11 & 12 & 13 \\ \hline \hline
\rotatebox{90}{\includegraphics[height=0.022cm]{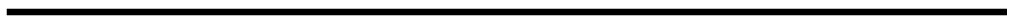}}
& 
\rotatebox{90}{\includegraphics[height=0.32cm]{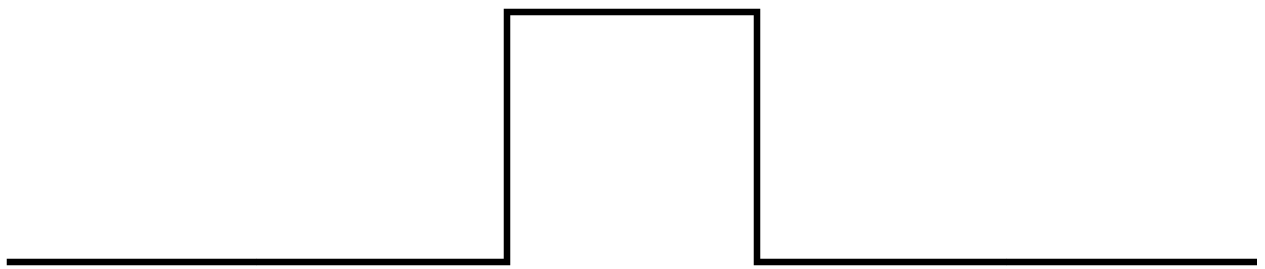}}
& 
\rotatebox{90}{\includegraphics[height=0.4cm]{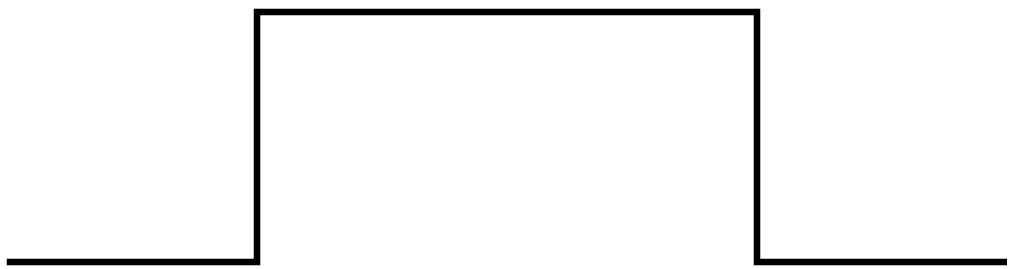}} 
& 
\rotatebox{90}{\includegraphics[height=0.8cm]{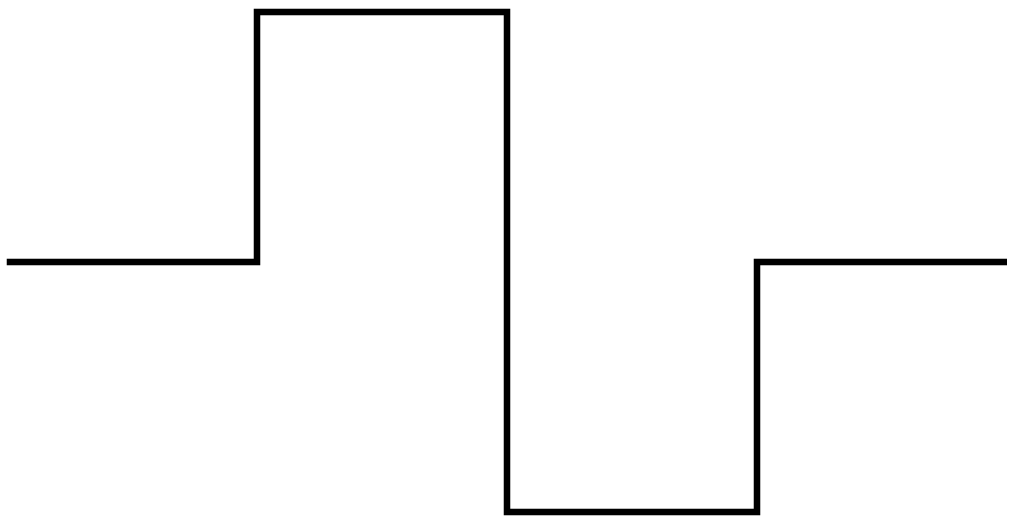}}
& 
\rotatebox{90}{\includegraphics[height=0.4cm]{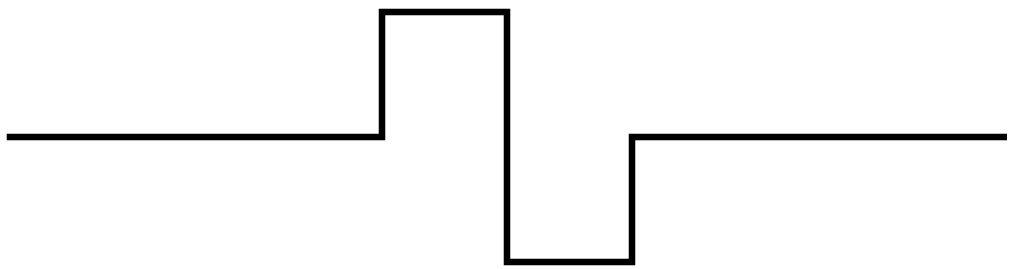}}
& 
\rotatebox{90}{\includegraphics[height=0.4cm]{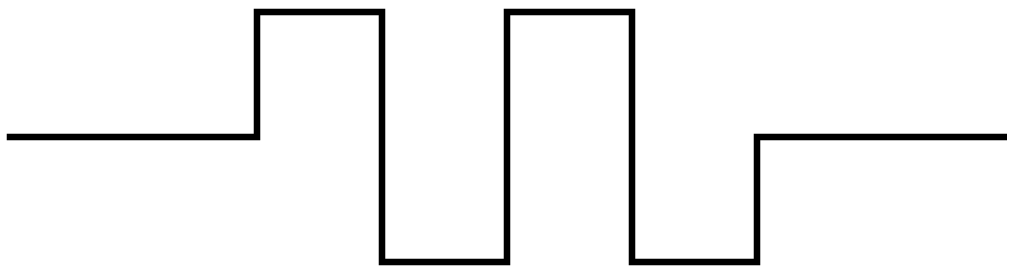}}
& 
\rotatebox{90}{\includegraphics[height=0.4cm]{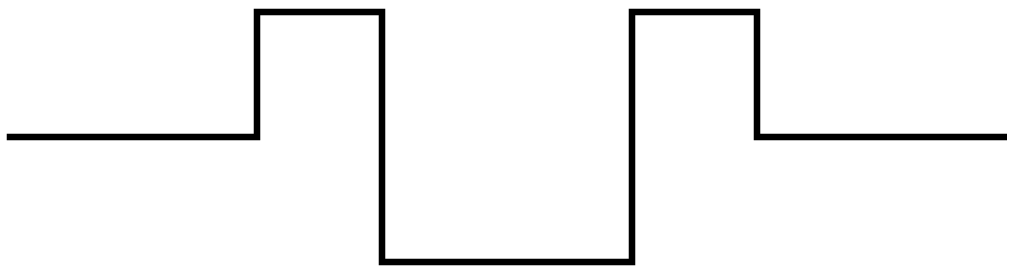}}
& 
\rotatebox{90}{\includegraphics[height=0.2cm]{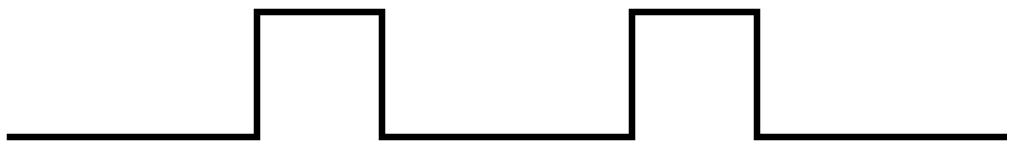}}
& 
\rotatebox{90}{\includegraphics[height=0.4cm]{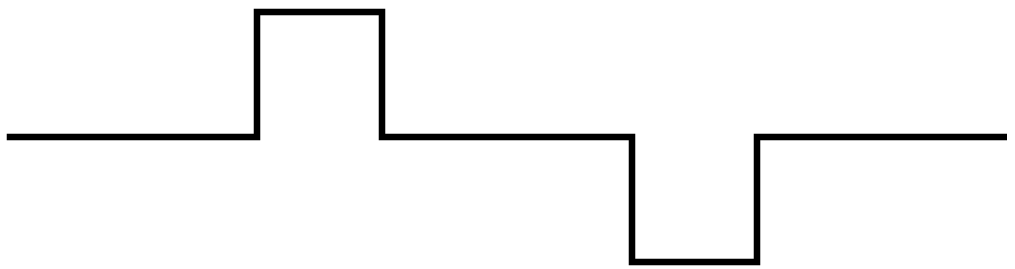}}
& 
\rotatebox{90}{\includegraphics[height=0.2cm]{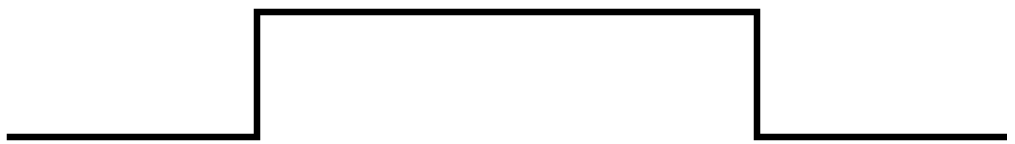}}
& 
\rotatebox{90}{\includegraphics[height=0.4cm]{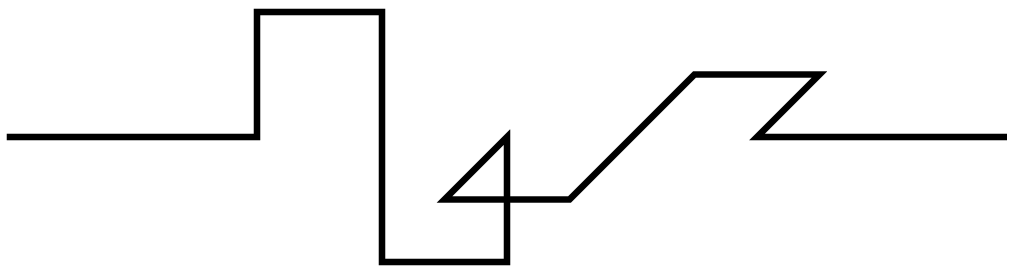}} 
& 
\rotatebox{90}{\includegraphics[height=0.32cm]{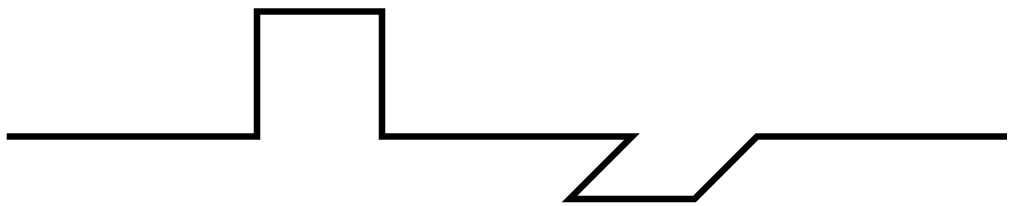}} 
& 
\rotatebox{90}{\includegraphics[height=0.32cm]{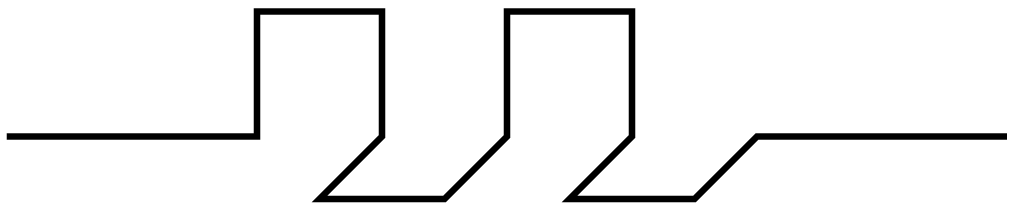}} \\ \hline \hline \hline
14&15&16&17&18&19&20&21&22&23&24&25&26 \\ \hline
\rotatebox{90}{\includegraphics[height=0.4cm]{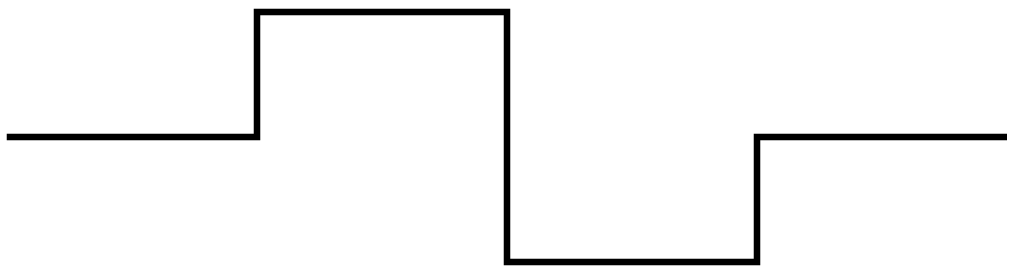}}
& 
\rotatebox{90}{\includegraphics[height=0.32cm]{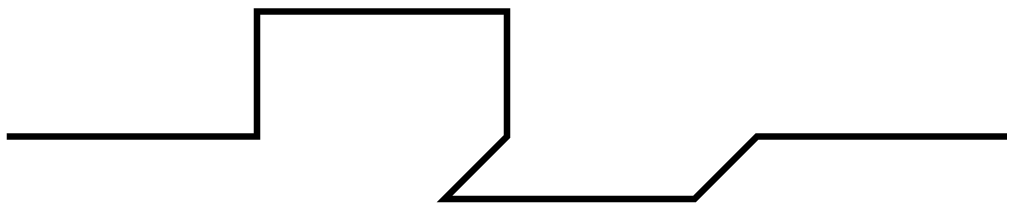}}
& 
\rotatebox{90}{\includegraphics[height=0.32cm]{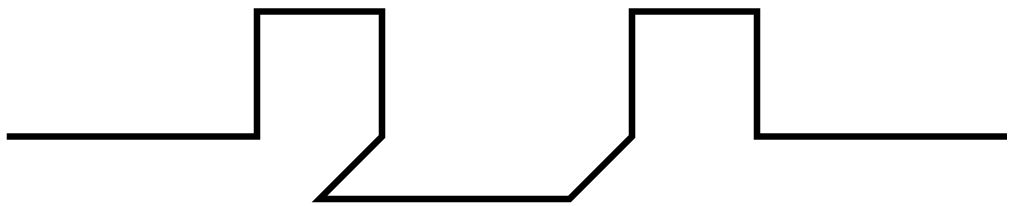}}
& 
\rotatebox{90}{\includegraphics[height=0.32cm]{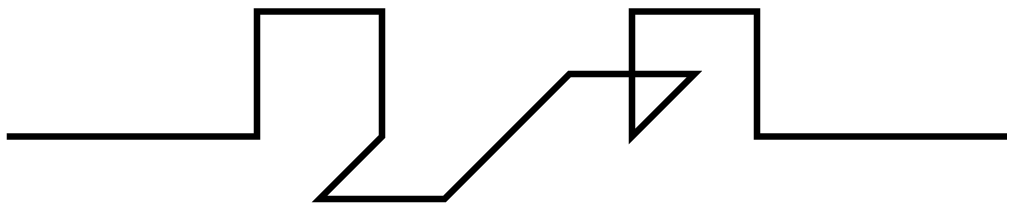}}
& 
\rotatebox{90}{\includegraphics[height=0.4cm]{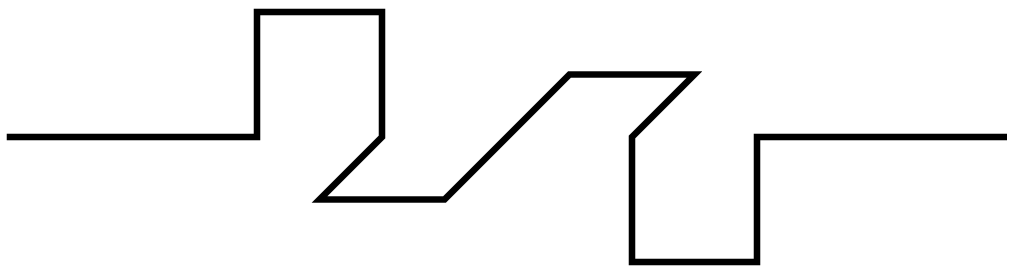}}
& 
\rotatebox{90}{\includegraphics[height=0.32cm]{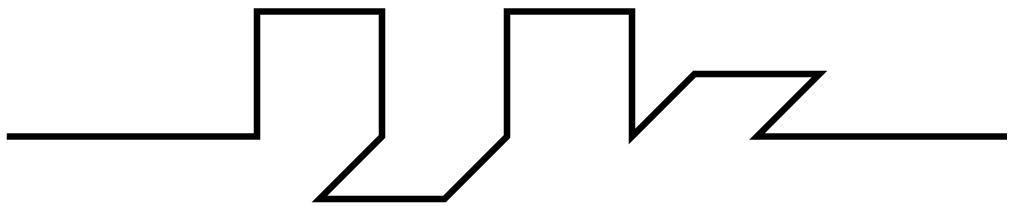}}
& 
\rotatebox{90}{\includegraphics[height=0.4cm]{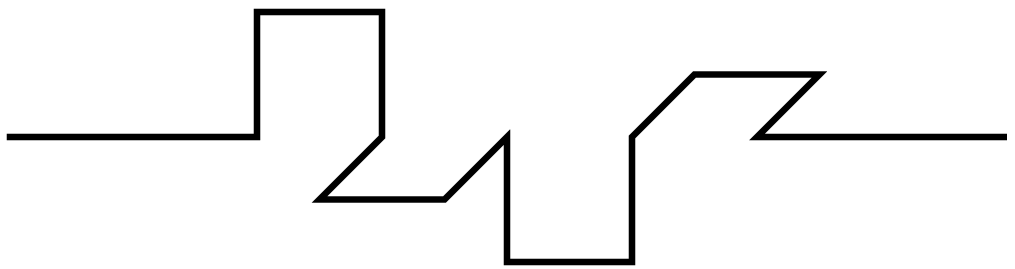}}
& 
\rotatebox{90}{\includegraphics[height=0.32cm]{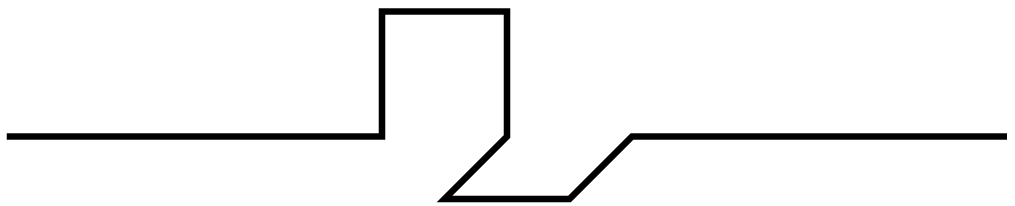}}
& 
\rotatebox{90}{\includegraphics[height=0.64cm]{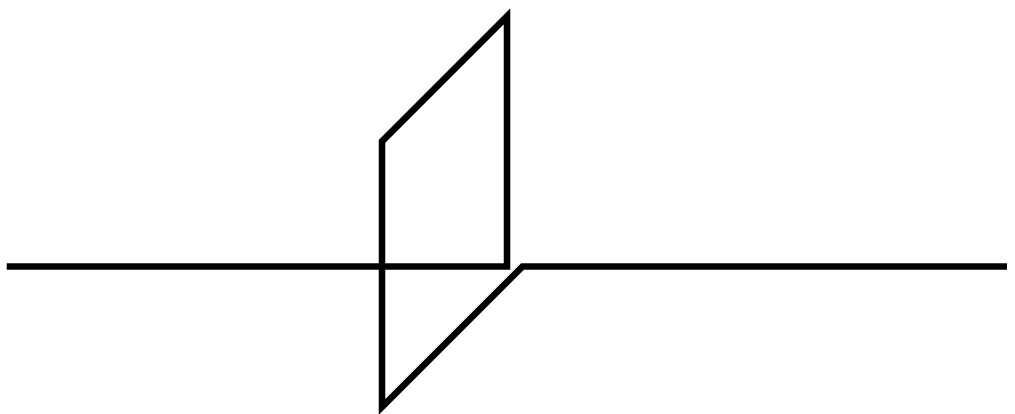}}
& 
\rotatebox{90}{\includegraphics[height=0.64cm]{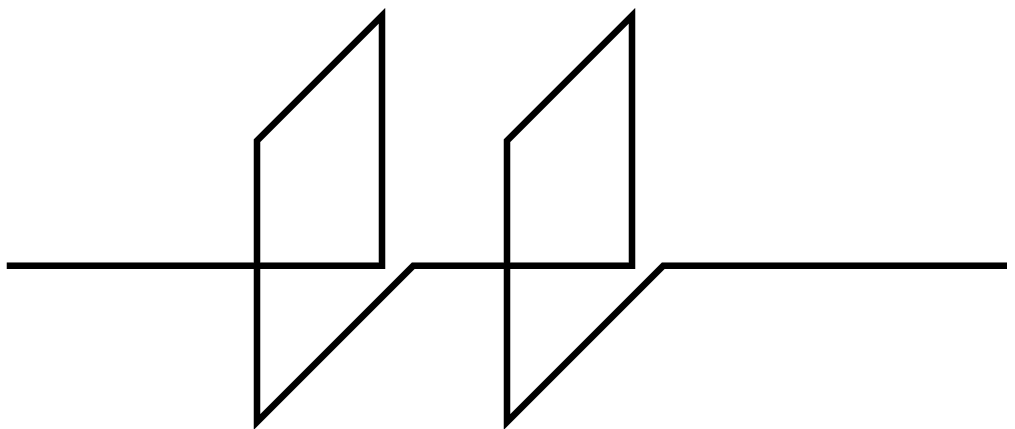}}
& 
\rotatebox{90}{\includegraphics[height=0.8cm]{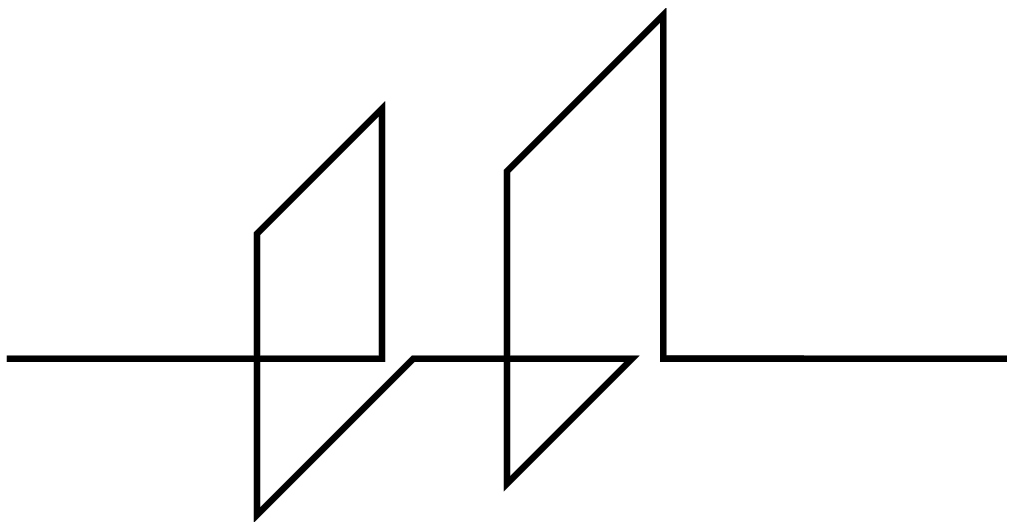}}
& 
\rotatebox{90}{\includegraphics[height=0.64cm]{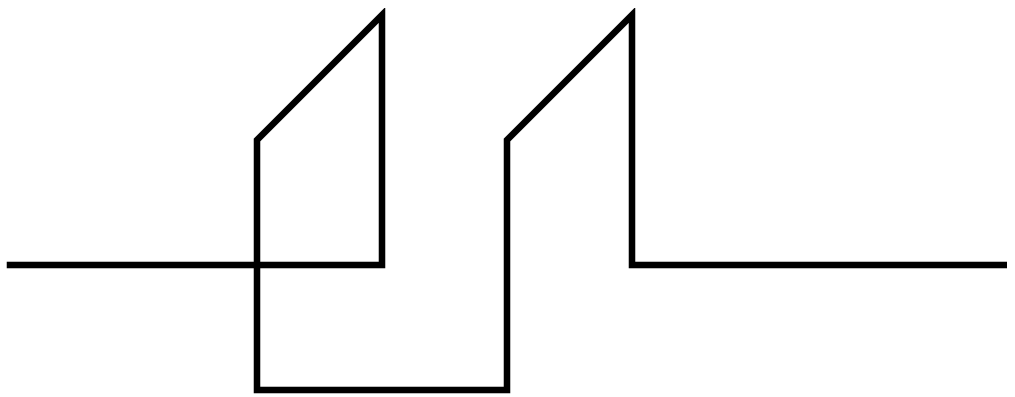}}
& 
\rotatebox{90}{\includegraphics[height=0.64cm]{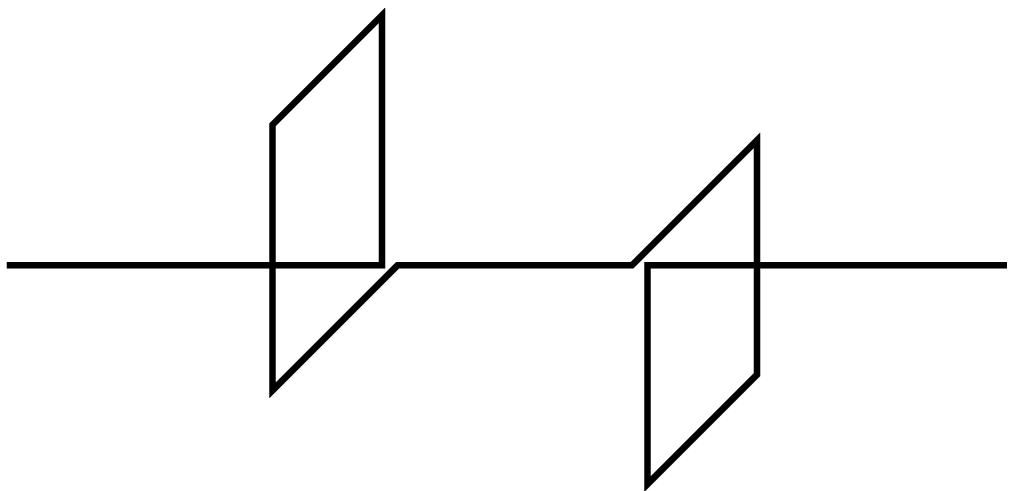}}\\ \hline
\end{tabular}}}
\caption{All the transverse deformations used for the construction of the operators.}
\label{operatorstable}
\end{table}
\vspace{-0.75cm}
\section{Theoretical expectations for the spectrum from effective string theories}
\label{theory}
\vspace{-0.2cm}
Let us first think about the flux-tube as a string of length $l=aL_{\parallel}$ winding around the torus. The classical configuration of the string spontaneously breaks translation symmetry and so we expect a set of Nambu-Goldstone massless bosons to appear at low energies. These bosons reflect the transverse fluctuations of the flux-tube around its classical configuration, and in $D$ space-time dimensions there are $D-2$ of them. Below, we describe the current theoretical predictions for the excitation spectrum of these bosons.
\vspace{-0.25cm}
\subsubsection{The Nambu-Goto model}
%\vspace{-0.25cm}
The Nambu-Goto model describes relativistic strings \cite{NGpapers}. The action of the model is proportional to  the area of the world sheet swept by the string as it propagates in time. This model is self-consistent quantum mechanically only in $D=26$ dimensions (see for example second reference in \cite{NGpapers}), but there are claims in the literature that, for any value of $D$, this model can serve as an effective low energy field theory for long strings \cite{Olesen}. From here on we refer to such low energy theories as Effective String Theories (ESTs).
The spectrum of the Nambu-Goto model for $D=4$ is given below (e.g.~see second reference in \Ref{NGpapers}).
%\vspace{-0.25cm}
\begin{eqnarray}
E_{\rm NG}(l)=\sqrt{(\sigma l)^2 +4 \pi \sigma \Big( N^+_L + N^-_L + N^+_R + N^-_R - \frac{1}{6}\Big)  + \left(\frac{2\pi q}{l}\right)^2}.
\label{energy}
\end{eqnarray}
%\vspace{-0.65cm}
Here $\sigma$ is the string tension and $N^{\pm}_{L}$ and $N^{\pm}_R$ are the occupation numbers of the bosons that move to the left and to the right (the $\pm$ superscript denotes the spin that they carry). This means that each of these occupation numbers is defined to count the energy units carried by the bosons: $N = \sum_{k=1}^\infty k \, n_k$ (here $n_k$ is the number of bosons carrying momentum $k$). The net longitudinal momentum carried by the bosons is given by $q=N^+_L + N^-_L - N^+_R -N^-_R$, and the net angular momentum of a state is given by $J=N^+_L + N^+_R - N^-_L - N^-_R$. It is useful to make a connection to Regge theory by writing $\sum_{i=\pm} \sum_{j=L,R} N^i_j\equiv J+m$ and then interpreting the integer $m$ as counting the daughter trajectories of a certain angular momentum $J$. As usual, leading Regge trajectories of angular momentum $J$ can be degenerate in energy with daughter trajectories of lower spin states. 
Below we refer to states that have the same value of $J+m$ as `being in the same NG level'.

Since we think of the NG model as an EST, which may be justified only for long strings \cite{Olesen}, we can expand \Eq{energy} for $l\surd \sigma\gg 1$ and we get (for simplicity we set $q=0$, and follow the convention in denoting $J+m$ by $n/2$) 
\begin{eqnarray}
  E_{\rm NG}(l) =  \sigma l + \frac{4 \pi}{l}\bigg(n-\frac{D-2}{24}\bigg)  -  \frac{8 \pi^2}{\sigma l^3}\bigg(n-\frac{D-2}{24}\bigg)^2  +   \frac{32 \pi^3}{\sigma^2 l^5}\bigg(n-\frac{D-2}{24}\bigg)^3   + {O}(l^{-7}).
\label{AharonyKarzbrun}
\end{eqnarray}
We note in passing that because the NG model is only one possible candidate of an EST, then in the language of effective field theories, it may differ from other candidate ESTs by the values of certain low energy constants (LECs). These LECs would make the most general EST spectrum differ from \Eq{AharonyKarzbrun} by the coefficients in the $1/l$ expansion. In the next subsection we discuss a systematically controlled approach to construct the most general EST of the QCD flux-tube.

\vspace{-0.25cm}
\subsubsection{Effective string theory approaches}
\label{effectivepred}
%\vspace{-0.25cm}
A systematic EST
 study that would describe the QCD flux-tube was pioneered by  L\"uscher, Symanzik, and Weisz in \Ref{Luscher}. Such an EST approach produces predictions for the energy of states as an expansion in $1/l$. Terms in this expansion that are of $O(1/l^p)$ are generated by $(p+1)$-derivative terms in the EST action whose coefficients are a priori arbitrary LECs. Interestingly, these LECs were shown to obey strong constraints that reflect a non-linear realization of Lorentz symmetry \cite{LW,Meyer,AK}, and so to give parameter free predictions for certain terms in the $1/l$ expansion. We review these predictions below.

First, since we focus on closed strings, $p$ can only be odd (terms in the energy that come with even powers of $1/l$ appear only if there are boundary terms in the action of the EST and so do not exist for closed strings). The analysis with $p=1$  was performed in \Ref{Luscher} and was shown to yield the $O(1/l)$ `L\"uscher term' in the ground state energy, whose universal coefficient depends only on $D$. Since the $2-$derivative action is a free theory, then its spectrum of excited states is that of $(D-2)$ massless bosons (the L\"uscher term is the zero point energy of these bosons). The predictions of \Ref{Luscher} were that to $O(1/l)$  the flux-tube energy is given by the first two terms in right hand-side of \Eq{AharonyKarzbrun}.

The $4-$derivative terms were analysed in \Ref{LW} and for $D=3$ were shown to yield an $O(1/l^3)$ term that is identical to the third term in \Eq{AharonyKarzbrun}. \Ref{AK} showed that this matching between the $O(1/l^3)$ term in the NG prediction and in any general EST holds also for $D=4$. \Ref{AK} also analyzed the $6-$derivative terms and showed that for $D=3$ they yield the fourth term in the r.h.s. of \Eq{AharonyKarzbrun}, while for general states in $D=4$,  the coefficient of the $O(1/l^5)$ term may differ from the one in \Eq{AharonyKarzbrun}. Nonetheless, the energy of $n=0$ state in the $D=4$ case is special and \Ref{AK} showed that its $O(1/l^5)$  term is indeed given by \Eq{AharonyKarzbrun}, and that the $O(1/l^5)$ term in the average over the energies of states that are in the same NG level is identical to the $O(1/l^5)$ in \Eq{AharonyKarzbrun}. 

A different approach to EST was proposed by Polchinski and Strominger in \Ref{Pol}. Technically it uses a different gauge fixing of the embedding coordinates on the world-sheet (conformal gauge instead of the static gauge choice used in the ESTs following \Ref{Luscher}). Here the constraints obeyed by the LECs allows one to maintain the conformal symmetry of the world-sheet even outside the critical dimension of $D=26$. \Ref{Pol} showed that as a result of these constraints, the  $O(1/l)$ term in the $1/l$ expansion is the same as in the NG model --- it is given by the Luscher term appearing in \Eq{AharonyKarzbrun}. Much more recently, Drummond \cite{Drummond} showed that the $O(1/l^3)$ term is also identical to the one appearing in the $1/l$ expansion of the NG model (\Eq{AharonyKarzbrun}). Finally, the recent Refs.\cite{DM} claim that even higher order terms in the $1/l$ expansion are identical to the corresponding terms in \Eq{AharonyKarzbrun}; these claims seem to  contradict the results of \Ref{AK} for $D\ge 4$.
\vspace{-0.45cm}
\section{Results}
\vspace{-0.3cm}
\subsection{Ground states in the $q=0,1,2$ channels}
\begin{figure}[htb]
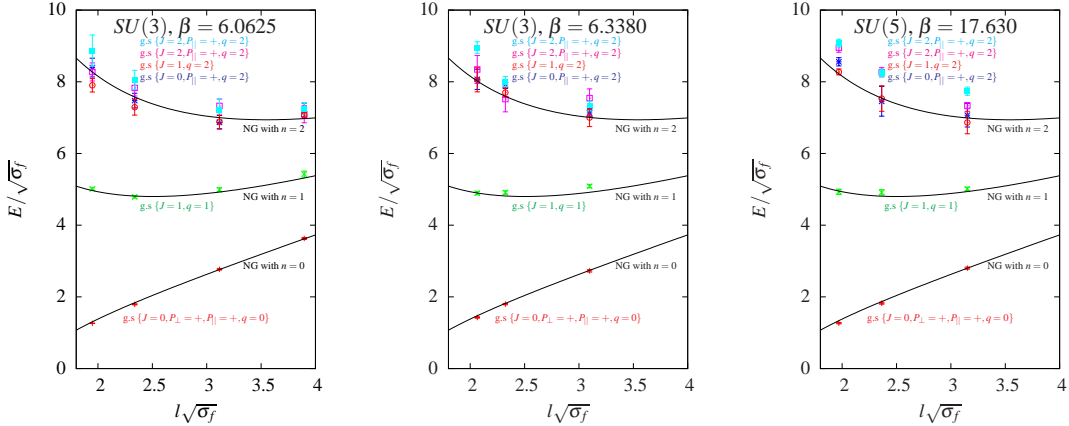

\centerline{\hspace{-00mm}
  \scalebox{0.65}{\input{plotproceedingsgroundSU3B6.06253.tex} \put(-130,90){\tiny NG with $n=0$} \put(-130,130){\tiny NG with $n=1$} \put(-130,170){\tiny NG with $n=2$} \put(-200,60){\Red{\tiny g.s $ \{J=0, P_\perp=+, P_{||}=+,q=0  \}$}} \put(-190,125){\Green{\tiny g.s $ \{J=1, q=1  \}$}} \put(-190,200){\Blue{\tiny g.s $ \{J=0, P_{||}=+, q=2  \}$}}  \put(-190,207){\Red{\tiny g.s $ \{J=1, q=2  \}$}}  \put(-190,214){\Magenta{\tiny g.s $ \{J=2, P_{||}=+, q=2  \}$}}  \put(-190,221){\Cyan{\tiny g.s $ \{J=2, P_{||}=+,q=2  \}$}}  } \hspace{-35mm}  \scalebox{0.65}{\input{plotproceedingsgroundSU3B6.338v2.tex} \put(-130,90){\tiny NG with $n=0$} \put(-130,130){\tiny NG with $n=1$} \put(-130,170){\tiny NG with $n=2$} \put(-200,60){\Red{\tiny g.s $ \{J=0, P_\perp=+, P_{||}=+,q=0  \}$}} \put(-190,125){\Green{\tiny g.s $ \{J=1, q=1  \}$}} \put(-190,200){\Blue{\tiny g.s $ \{J=0, P_{||}=+, q=2  \}$}}  \put(-190,207){\Red{\tiny g.s $ \{J=1, q=2  \}$}}  \put(-190,214){\Magenta{\tiny g.s $ \{J=2, P_{||}=+, q=2  \}$}}  \put(-190,221){\Cyan{\tiny g.s $ \{J=2, P_{||}=+,q=2  \}$}}    } \hspace{-35mm}
\scalebox{0.65}{\input{plotproceedingsgroundSU5B17.630.tex} \put(-130,90){\tiny NG with $n=0$} \put(-130,130){\tiny NG with $n=1$} \put(-130,170){\tiny NG with $n=2$} \put(-200,60){\Red{\tiny g.s $ \{J=0, P_\perp=+, P_{||}=+,q=0  \}$}} \put(-190,125){\Green{\tiny g.s $ \{J=1, q=1  \}$}} \put(-188,200){\Blue{\tiny g.s $ \{J=0, P_{||}=+, q=2  \}$}}  \put(-188,207){\Red{\tiny g.s $ \{J=1, q=2  \}$}}  \put(-188,214){\Magenta{\tiny g.s $ \{J=2, P_{||}=+, q=2  \}$}}  \put(-188,221){\Cyan{\tiny g.s $ \{J=2, P_{||}=+,q=2  \}$}} }
\put(-415,148){\scriptsize $SU(3)$, $\beta=6.0625$} \put(-275,148){\scriptsize $SU(3)$, $\beta=6.3380$} \put(-135,148){\scriptsize $SU(5)$, $\beta=17.630$}}
\vspace{-0.3cm}
\caption{Energies of the lightest states that correspond to $q=0,1,2$ NG ground states.}% for $SU(3)$ at $\beta=6.0625$, $SU(3)$ at $\beta=6.3380$ and $SU(5)$ at $\beta=17.630$.}
\label{3+1dground}
\vspace{-0.7cm}
\end{figure}

\vspace{-0.2cm}
In Figure~\ref{3+1dground} we present a comparison of the $q=0,1,2$ NG ground states with our data: $SU(3)$ and $a\simeq 0.09$fm in the left most plot, $SU(3)$ and $a\simeq 0.06$fm in the central plot, and $SU(5)$ and $a\simeq 0.09$fm in the right most plot. The NG ground 
state for $q=0$ is the state of no bosons, and has trivial quantum numbers $\{ J=0, P_\perp=+,P_{||}=+ \}$. Indeed our data shows that the ground state is in the $A_1$ representation with $q=0$ and $P_{||}=+$. We use the measured energy of this state to extract the string tension by fitting it with the ansatz $
E_{gs}(l) = \sqrt{\left(\sigma l\right)^2 - \frac{2\pi\sigma}{3} + \sigma \frac{C}{(\surd \sigma l)^{6}}}$; we find $C\sim{O}(1-10)$ and that for $l\sqrt{\sigma}\gtrsim 3$ the correction term is negligible. 

The NG ground state in the $q=1$ channel has $J=1$ and is $P_\perp$ degenerate, and so we measure only the energy of the $P_\perp=+$ channel in the $E$ representation. This measurement is presented in Fig.~\ref{3+1dground} where we see that the NG prediction is in agreement with our data. Finally, NG predicts that the ground state for $q=2$ should be five-fold degenerate, consisting of a state with $\{ J=0,P_\perp=+ \}$, two states with $\{ J=1,P_\perp=\pm \}$, and two states with $\{ J=2,P_\perp=\pm \}$. We find this to be consistent with the ground states in the $A_1,E$ and $B_{1,2}$ representation (again, as for $q=1$, we measure only the $P_\perp=+$ states in the $E$ representation). Comparing the three plots in Fig.~\ref{3+1dground} we see that the $O(a^2)$ and $O(1/N^2)$ corrections of our data are small compared to our statistical errors. Thus, within our  level of accuracy, the agreement of our data with the NG model is largely  insensitive to the lattice spacing and $1/N^2$ corrections.

%\vspace{-0.45cm}
\vspace{-0.2cm}
\subsection{Further $q=0$ states}
\begin{figure}[htb]
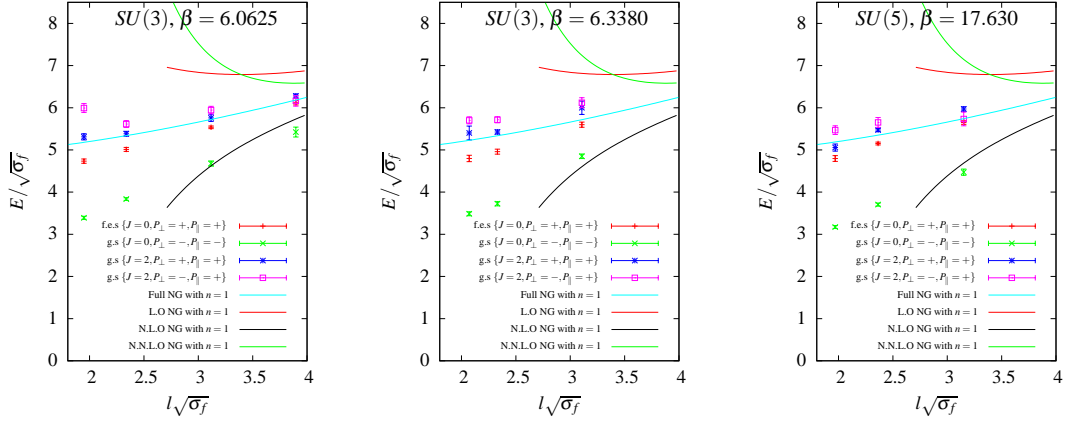

%\vspace{-0.55cm}
\centerline{\hspace{-00mm}
  \scalebox{0.65}{\input{plotproceedingsfirstq0SU3B6.06252.tex}} \hspace{-35mm}  \scalebox{0.65}{\input{plotproceedingsfirstq0SU3B6.338.tex} } \hspace{-35mm}
\scalebox{0.65}{\input{plotproceedingsfirstq0SU5B17.630.tex}}
\put(-415,148){\scriptsize $SU(3)$, $\beta=6.0625$} \put(-275,148){\scriptsize $SU(3)$, $\beta=6.3380$} \put(-135,148){\scriptsize $SU(5)$, $\beta=17.630$}}
%\vspace{-0.25cm}
\caption{Energies of the four states with $q=0$ expected to be characterised by $n=1$.}%, for $SU(3)$ at $\beta=6.0625$, $SU(3)$ at $\beta=6.3380$ and $SU(5)$ at $\beta=17.630$.%Lines correspond to the NG predictions.% for $D=4$ and the string tension obtained by fitting the ground state $\{ A_1,P_{||}=+ \}$ (or $\{ J=0, P_\perp=+, P_{||}=+ \}$).
%}
\label{3+1firstq0}
%\vspace{-0.55cm}
\end{figure}

\vspace{-0.2cm}
We now turn to examine the first excited state in the $q=0$ sector and present the results in Figure~\ref{3+1firstq0}. In the  NG model, this energy level is four-fold degenerate with a single state from each of the following representations: $\{ J=0,P_\perp=P_{||}=+ \}$, $\{ J=0,P_\perp=P_{||}=- \}$, $\{ J=2,P_\perp=P_{||}=+\}$, and  $\{ J=2,P_\perp=-,P_{||}=+ \}$. On the lattice, this would imply that states from $(A_1,P_{||}=+),(A_2,P_{||}=-),(B_1,P_{||}=+)$, and $(B_2,P_{||}=+)$ are all degenerate (up to $O(a^2)$ corrections). 

Instead, what we find is that while the states belonging to $(A_1,P_{||}=+),(B_1,P_{||}=+)$, and $(B_2,P_{||}=+)$ are all quite close to each other and to the NG model, the state in $(A_2,P_{||}=-)$ is anomalously different and shows substantial deviation from NG. It is tempting to expect that this state's energy would eventually approach the NG prediction (and is perhaps reflecting a large coefficient multiplying the term that controls this deviation from NG in the EST). Nonetheless, an equally likely possibility is that it does not converge to NG. In fact, observe that the energy of this state is higher than the ground state energy by approximately an equal amount throughout the distance range that our simulations are able to explore. Therefore, at large enough $l$, it might cross the NG prediction, as a massive state would. We do not know which possibility provides a better explanation for the strikingly `anomalous' way that the energy of this state behaves. This behaviour does not change as we decrease the lattice spacing (central plot) or increase $N$ (right-most plot). 

In Figure~\ref{3+1firstq0} we also provide a comparison of the four states with expansions of the Nambu-Goto square root order by order in terms of $1/l$ up to $O(1/l^5)$ -- see \Eq{AharonyKarzbrun} (note that according to \Ref{AK} it is only up to $O(1/l^3)$ that we can trust \Eq{AharonyKarzbrun}). Excluding the anomalously behaving ground state in the $\{ A_2, P_{||}=- \}$ channel, the other three states are obviously better described by NG than by any other EST. This reflects a simple fact: nearly all our data is beyond the radius of convergence of the $1/l$ expansion (which  can be estimated from the EST series to be  $\left(l\surd\sigma\right)_{\rm converge} \simeq 3.5$ for this particular state). Thus, the fact the our data is so close to NG even beyond the radius of convergence suggests that the EST approach can be somehow resummed to yield the NG square root of \Eq{energy} plus some small corrections.
In our $2+1$ study \cite{ABM} we also saw an agreement with NG beyond the radius of convergence of the EST.
\vspace{-0.5cm}
\subsection{Further $q=1$ states}
\begin{figure}[htb]
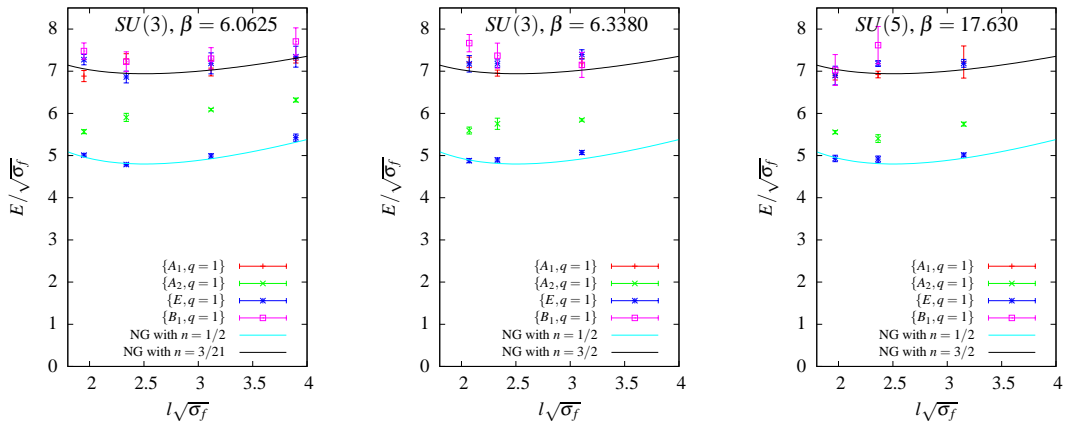

%\vspace{-0.55cm}
\centerline{\hspace{-00mm}
  \scalebox{0.65}{\input{plotproceedingsfirstSU3B6.0625.tex}} \hspace{-35mm}  \scalebox{0.65}{\input{plotproceedingsfirstSU3B6.338.tex} } \hspace{-35mm}
\scalebox{0.65}{\input{plotproceedingsfirstSU5B17.630.tex}}
\put(-415,148){\scriptsize $SU(3)$, $\beta=6.0625$} \put(-275,148){\scriptsize $SU(3)$, $\beta=6.3380$} \put(-135,148){\scriptsize $SU(5)$, $\beta=17.630$}}
%\vspace{-0.25cm}
\caption{Energies of the lightest five distinguishable states with $q=1$.}% for $SU(3)$ at $\beta=6.0625$, $SU(3)$ at $\beta=6.3380$ and $SU(5)$ at $\beta=17.630$. %Lines correspond to the NG predictions.% for $D=4$ and the string tension obtained by fitting the ground state $\{ A_1,P_{||}=+ \}$ (or $\{ J=0, P_\perp=+, P_{||}=+ \}$).
%}
\label{3+1firstq1}
\vspace{-0.5cm}
\end{figure}

\vspace{-0.17cm}
We now discuss further results for states with $q=1$ --- see Figure~\ref{3+1firstq1}. Here the first excited NG energy level ($n = 3/2$) should be ten-fold degenerate. These states are the ground states of $\{ J=0,P_\perp=+ \}$, $\{ J=0,P_\perp=- \}$, $\{ J=3, P_\perp=\pm \}$, $\{ J=2,P_\perp=+ \}$, $\{ J=2,P_\perp=- \}$ and the first  excited states of $\{ J=1, P_\perp=\pm \}$. On the lattice these states fall into two degenerate pairs of states in $E$, and four more states that belong to $A_{1,2}$ and $B_{1,2}$. The parity degeneracy in the $E$ representation allows us to calculate only the $P_\perp=+$ states and we therefore expect six degenerate states in the NG model. 

Unfortunately, our basis of operators was insufficient to successfully isolate all these six states, and we were able to extract only four of them. We found that the energy of the $q=1$ lowest energy states in the $A_1$ and $B_1$ representation, and second lowest energy states in the 
$E$ representation, agree fairly well with the energy of the $n=3/2$ NG level. In contrast, the ground state of $q=1$ in the  $A_2$ representation (which is naively associated with the $J=0$, $P_\perp=-$ channel) appears  anomalous: it has a large  deviation from the NG curve and does not show any sign of convergence. This is true also on our finer lattice (central plot) and for $SU(5)$ (right-most plot).
%\subsection{Comparison to Effective Strings}
\vspace{-0.75cm}
\section{Summary}
\vspace{-0.45cm}
We calculated energies of  13 states in the spectrum of closed strings in $3+1$ dimensional $SU(N)$ gauge theories. We study the gauge groups $SU(3)$ (with lattice spacing $a\approx 0.09,0.06$fm) and $SU(5)$ (with $a\approx 0.09$fm) and the string lengths in the range $0.9\,{\rm fm}\lesssim l \lesssim 1.8\, {\rm fm}$.

Most of the energies we measured show a convincing agreement with the Nambu-Goto (NG) model. We compare our results with recent predictions from other effective string theories (ESTs), but because most of the flux-tubes that we probe have lengths that are beyond the radius of convergence of the ESTs (for the excited states), this comparison fails. The only model that describes our data well, even for short flux-tubes and excited states, is the NG model. This was seen also in our previous work on $2+1$ dimensions (see \Ref{ABM}), and seems to suggest that the EST expansion can be resummed. Despite the overall good agreement with NG, we do find large deviations for certain states that have negative parity and that belong to the $A_1$ representation of the lattice group (which naively corresponds to zero angular momentum). We see these deviations also for states that carry both zero and one unit of longitudinal momentum. 

There are many avenues one could take on the lattice to make progress towards establishing what is the effective string theory of the QCD flux-tube. These future studies may include the search for massive excitations like breathing modes, attempting to accurately test the current theoretical predictions (see \Sec{theory}) within their radius of convergence, studying the open string spectrum, etc. We also look forward to theoretical progress that would allow one to understand the results we presented in this proceedings. For example, how can the flux-tube behave like a NG string below the radius of convergence of the effective string theory expansion? and what makes the states in the $A_1$ representation have large deviations from the NG model?

\vspace{-0.6cm}
\section*{Acknowledgements}
\vskip -0.4cm
BB thanks O.~Aharony for multiple useful discussions and the Weizmann institute for its kind hospitality. The computations were carried out on EPSRC and Oxford funded computers in Oxford Theoretical Physics. 
AA acknowledges the support of the Leventis Foundation and the LATTICE 2009 organizers for supporting him financially. 
BB was supported by the U.S. Department of Energy under Grant No. DE-FG02-96ER40956.

\end{document}